\documentclass{cta-author}

{}
{}
{}
\usepackage{subfigure}
\usepackage{tabularx}
\usepackage{multirow}
\usepackage{makecell}
\usepackage{color}
\usepackage{hhline}

\begin{document}

\supertitle{Submission Template for IET Research Journal Papers}

\title{Cuboid-Net: A Multi-Branch Convolutional Neural Network for Joint Space-Time Video Super Resolution}

\author{\au{Congrui Fu$^{1}$}, \au{Hui Yuan$^{2\corr}$}, \au{Hongji Xu$^{1}$}, \au{Hao Zhang$^{3}$}, \au{Liquan Shen$^{4}$}}

\address{\add{1}{School of Information Science and Engineering, Shandong University, Qingdao, China}
\add{2}{School of Control Science and Engineering, Shandong University, Jinan, China}
\add{3}{School of Information and Control Engineering, Qingdao University of Technology, Qingdao, China}
\add{4}{School of Communication and Information Engineering, Shanghai University, Shanghai, China}
\email{huiyuan@sdu.edu.cn}}

\begin{abstract}
The demand for high-resolution videos has been consistently rising across various domains, propelled by continuous advancements in science, technology, and societal.
Nonetheless, challenges arising from limitations in imaging equipment capabilities, imaging conditions, as well as economic and temporal factors often result in obtaining low-resolution images in particular situations.
Space-time video super-resolution aims to enhance the spatial and temporal resolutions of low-resolution and low-frame-rate videos. 
The currently available space-time video super-resolution methods often fail to fully exploit the abundant information existing within the spatio-temporal domain.
To address this problem, we tackle the issue by conceptualizing the input low-resolution video as a cuboid structure. 
Drawing on this perspective, we introduce an innovative methodology called "Cuboid-Net," which incorporates a multi-branch convolutional neural network.
Cuboid-Net is designed to collectively enhance the spatial and temporal resolutions of videos, enabling the extraction of rich and meaningful information across both spatial and temporal dimensions.
Specifically, we take the input video as a cuboid to generate different directional slices as input for different branches of the network. The proposed network contains four modules, i.e., a multi-branch-based hybrid feature extraction (MBFE) module, a multi-branch-based reconstruction (MBR) module, a first stage quality enhancement (QE) module, and a second stage cross frame quality enhancement (CFQE) module for interpolated frames only. Experimental results demonstrate that the proposed method is not only effective for spatial and temporal super-resolution of video but also for spatial and angular super-resolution of light field. 
\end{abstract}

\maketitle

\section{Introduction}\label{sec1}

Nowadays, high definition videos are becoming more and more popular in people’s daily lives. However, due to the limitations of video acquisition devices and video transmission systems, videos with high spatial and temporal resolutions are usually hard to be sent to users with low cost. Therefore, video super resolution (VSR) becomes a critical technology to address this problem.

\begin{figure}[!h]
	\centering
	\includegraphics [width=\linewidth]{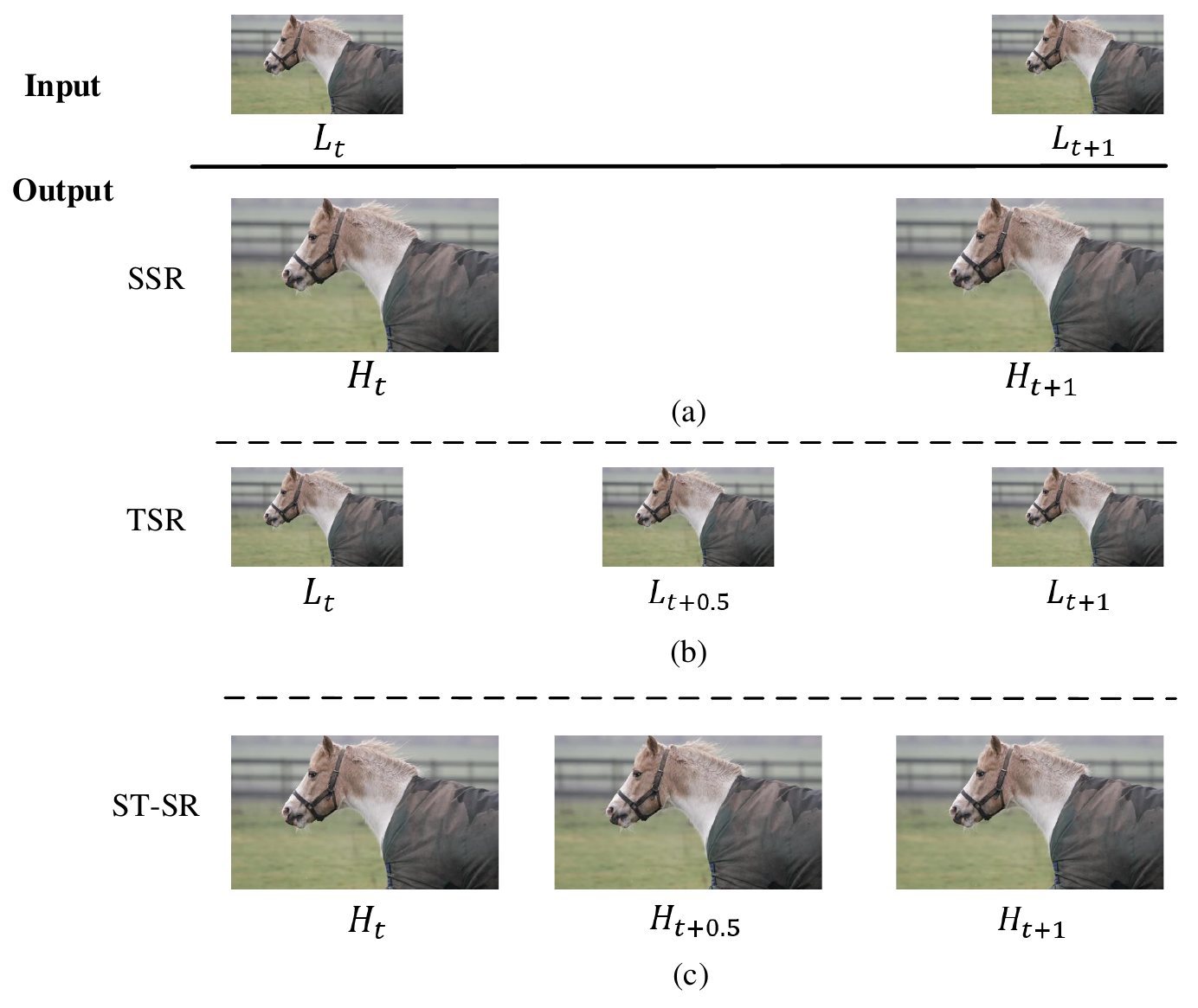}
	%%\centerline {\includegraphics [scale=0.5]{fig1.pdf}}
	\caption{Video super resolution, (a) spatial super resolution, (b) temporal super resolution, (c) space-time super resolution. $ H $ and $ L $ denote the high and low spatial resolution respectively, the subscripts $ t $, $ t+1 $, and $ t+0.5 $ denotes the time index.}
	\label{VSR}	
\end{figure}

There are three kinds of methods for VSR, i.e., spatial super resolution (SSR), temporal super resolution (TSR), and space-time super resolution (ST-SR). SSR is to increase the spatial size of each frame for a given video, while TSR is to interpolate/synthesize the uncollected frames of the video, as shown in Fig.~\ref{VSR}(a) and (b), respectively. The ST-SR is to automatically generate a photo-realistic video with high space-time resolution from a given video with low space-time resolution, as shown in Fig.~\ref{VSR}(c). In this paper, ST-SR is focused and VSR refers to ST-SR in the following. 

Existing ST-SR methods usually treat spatial and temporal super-resolution independently. The problem is that TSR and SSR have to be performed alternatively to obtain super-resolved videos, which is time consuming. Besides, when performing SSR first, it is hard to benefit from the high temporal resolution, and when performing TSR first, it is also hard to benefit from the high spatial resolution. Actually, SSR and TSR can benefit from each other. Higher spatial resolution can help to improve the motion estimation accuracy, which is important for TSR, while higher temporal resolution can also provide more accurate high-frequency details for SSR due to the similar appearance of successive frames.

Recently, as convolutional neural network (CNN) has proven to be efficient for image processing and computer vision tasks, great achievements for VSR have also been made by CNN (see Section II for details). The basic concept of CNN-based VSR is to efficiently extract the spatial and temporal features of an input video and reconstruct the output video with high spatial and temporal resolution by fully exploiting these features.

\begin{figure*}[!h]
	\centering{\includegraphics [width=\linewidth]{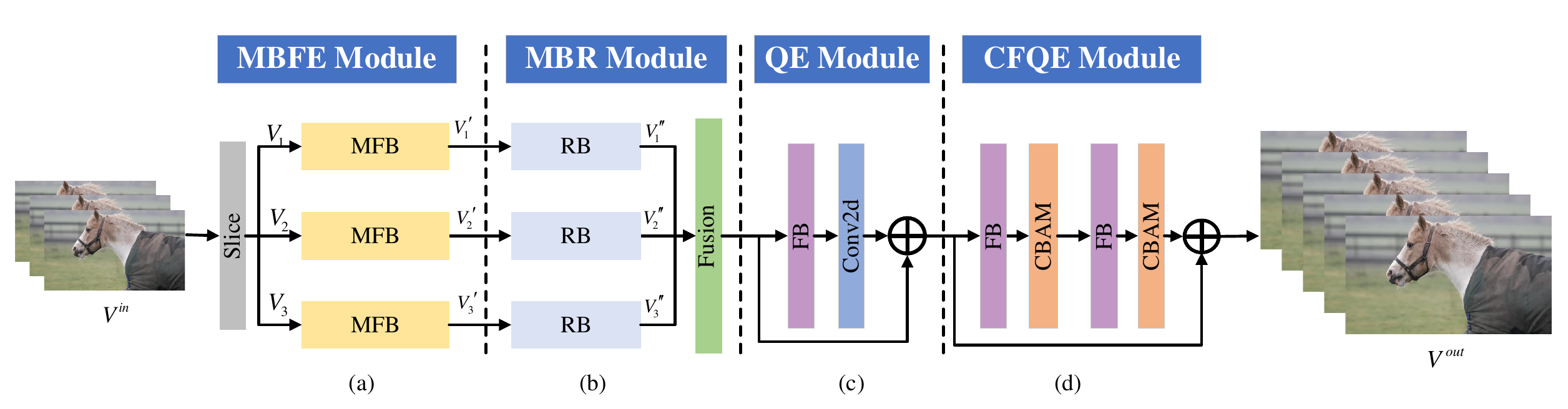}}
	\caption{Overview of the proposed Cuboid-Net.}
	\label{Net}
\end{figure*}

To make full use of the spatial and temporal features for VSR, we propose to generate the super-resolved video in both spatial and temporal domains simultaneously. Specifically, we propose a multi-branch convolutional neural network to efficiently extract hybrid spatial and temporal features for reconstruction. The proposed neural network consists of four modules: multi-branch-based hybrid feature extraction (MBFE) module, multi-branch-based reconstruction (MBR) module, a first stage quality enhancement (QE) module, and a second stage cross frame quality enhancement (CFQE) module for interpolated frames only, as shown in Fig.~\ref{Net}. 
The MBFE and MBR modules are used to extract hybrid features for spatial and temporal reconstruction, while the following QE and CFQE modules are used to further improve the reconstruction quality frame by frame. The contributions of this paper are as follows.

\begin{itemize}
	\item[$\bullet$] We propose a multi-branch-based hybrid feature extraction and reconstruction neural network that makes full use of the spatial and temporal information of successive frames to simultaneously generate the high spatial resolution and high frame rate video from a given input video. 
	
	\item[$\bullet$] By taking the input video as a cuboid, we propose to generate slices as inputs for different branches of the network by cutting the cuboid along with horizontal, vertical, and temporal directions, respectively, to efficiently extract and fully utilize the space-time features. Therefore, we call the proposed network Cuboid-Net.
	
	\item[$\bullet$] The proposed network structure can also be applied to the spatial and angular super resolution for other volumetric image data, e.g. light field, and also shows efficient performance.

\end{itemize}

The remainder of this paper is organized as follows. Related work on SSR, TSR, and ST-SR for videos are presented in Section II. The proposed network is then described in detail in Section III. Experimental results with ablation study are given in Section IV, which also extends the application of the proposed method to light field. Finally, Section V concludes the paper.

\section{Related work}\label{sec2}

\subsection{SSR for video}\label{subsec2.1}
SSR is to reconstruct a high-resolution video frame from its corresponding low resolution frames~\cite{1}.
Video SSR has evolved from its origins in image super-resolution techniques. The pioneering work in image super-resolution using Convolutional Neural Networks (CNN) is SRCNN~\cite{SRCNN}. 
Subsequently, numerous methods have arisen to further improve image super-resolution.
For instance, Liu \textit{et al.}~\cite{s1} introduced a network with self-attention negative feedback to attain real-time image super-resolution. Their network model effectively reduces the image mapping space and extracts crucial image features using the self-attention negative feedback mechanism. 
Shen \textit{et al.}~\cite{s2} proposed a deep learning-based algorithm for enhancing images in video surveillance contexts, employing a hybrid deep convolutional network. This approach utilizes a distinctive combination of morphological techniques and region merging through watershed-based enhancement.
Liu \textit{et al.}~\cite{s3} proposed a sharpened spatial filter image enhancement method for small targets. This method enhances small targets within subtle regions by amplifying the contrast between the object's edge and the surrounding image elements, thus enhancing the distinction of small targets from the background.
Different from single image super resolution~\cite{1}, SSR for video can additionally use the motion information of successive frames to compensate the lost high frequency details of the current frame. 

Liu and Sun~\cite{2} proposed a Bayesian approach to simultaneously estimate the underlying motion, blur kernel, and noise level to reconstruct the high resolution frame. 
Ma \textit{et al.}~\cite{3} tackled ubiquitous motion blurs by optimally searching the least blurred pixels for multi-frame super resolution. With the help of CNN, the mapping between low resolution frames and high resolution frames can be learned efficiently. 
Haris \textit{et al.}~\cite{4} proposed to perform SSR for video by incorporating CNN-based single image super resolution and optical flow estimation. Due to the difficulty of obtaining accurate optical flow, motion warping artifacts occur in the super-resolved frames. 
Tian \textit{et al.}~\cite{5} proposed a temporal deformable alignment network (TDAN) to adaptively align the reference frame and each supporting frame at feature level without optical flow estimation. This method implicitly captures motion cues via a deformable sampling module at the feature level and directly predicts aligned low resolution frames from sampled features without image-wise wrapping operations, and is capable of exploring image contextual information. 
Wang \textit{et al.}~\cite{6} proposed a video restoration framework with enhanced deformable convolutions, namely EDVR. The EDVR consists of pyramid, cascading, and deformable (PCD) alignment module to handle large motions and temporal and spatial attention (TSA) fusion module to emphasize important features for subsequent restoration. 
Li \textit{et al.}~\cite{7} proposed a deep dual attention network (DDAN) in which a motion compensation module and a super resolution reconstruction module are included to fully exploit the space-time informative features for accurate video SSR. 
Wang \textit{et al.}~\cite{8} proposed a video SSR network in which both the optical flow maps and frames are super-resolved. 
Li \textit{et al.}~\cite{43} proposed a temporal multi-correspondence aggregation strategy to leverage similar patches across frames, and a cross-scale nonlocal-correspondence aggregation scheme to explore self-similarity of images across scales. 
Chan \textit{et al.}~\cite{44} proposed a succinct pipeline for video SSR, namely BasicVSR, which uses bidirectional propagation to maximize information gathering, and an optical flow-based method to estimate the correspondence between two neighboring frames for feature alignment. 
Isobe \textit{et al.}~\cite{45} proposed to explore the role of explicit temporal difference modeling in both LR and HR space. Besides, recurrent neural networks~\cite{9}, such as convolutional LSTMs (ConvLSTM) are also adopted in video SSR methods~\cite{10,11} to efficiently exploit the temporal correlation among successive frames. 
Zhang \textit{et al.}~\cite{SSR2} proposed to extract spatio-temporal information from the aligned features and developed a shift-based deformable 3D convolution using low-cost bit shifts and additions.
Feng \textit{et al.}~\cite{SSR1} reconstructed a more real high-resolution high-frame-rate video using a hybrid video input, including a low-resolution high-frame-rate video (main video) and a high-resolution low-frame-rate video (auxiliary video).

\subsection{TSR for video}\label{subsec2.2}

TSR aims to generate an intermediate frame between any two given frames which is the same as video frame interpolation (VFI). It can be applied to numerous applications such as slow-motion generation, video frame rate conversion, virtual view synthesis, and frame recovery in video streaming~\cite{12,13}. In general, TSR methods are divided into two steps: motion estimation and intermediate frame synthesis. First, motion estimation which refers to measuring the direction and velocity of moving objects between two consecutive frames should be conducted. Based on the motion information, the intermediate frames can be interpolated along with the moving directions of each object. Second, the motion information and input frames are used to generate the pixel of the intermediate frame.

Mahajan \textit{et al.}~\cite{14} developed a moving gradient method that estimates paths in input images, copies proper gradients to each pixel in the frame to be interpolated and then synthesizes the intermediate frame via Poisson reconstruction. 
Meyer \textit{et al.}~\cite{15} proposed to propagate phase information of frequency domain across oriented multi-scale pyramid levels for VFI. Although these methods achieve impressive performance, the interpolation quality for high-frequency details with large motion is still not satisfactory. With the help of CNN, recently, the performance of VFI has been achieved significantly. 
Niklaus \textit{et al.}~\cite{16} took the frame interpolation problem as a local convolution problem over two successive frames and proposed to learn a spatially adaptive convolution kernel for each pixel by CNN. To reduce the computationally complexity, they further improved the efficiency by learning separable kernels. 
Liu \textit{et al.}~\cite{17} proposed a deep voxel flow (DVF) network to explorer the benefits of conventional optical-flow-based methods and deep learning-based methods. The basic idea of this method is to reconstruct pixels of the interpolated frames by using features of a 3D voxel flow vector across time and space of some input key frames, with trilinear sampling over the volume of the input video. 
Jiang \textit{et al.}~\cite{18} proposed an end-to-end CNN for variable-length multi-frame video interpolation. They computed bidirectional optical flow between the input frames by U-Net~\cite{19}, and employed another U-Net to refine the optical flow to predict soft visibility maps. Then the input frames were warped and linearly fused to reconstruct the intermediate frames. 
Bao \textit{et al.}~\cite{20} proposed an end-to-end network in which the input frames are first warped by the optical flow and then sampled via the learned interpolation kernels within an adaptive warping layer. 
Bao \textit{et al.}~\cite{21} also proposed another VFI method through depth-aware flow projection layer to synthesize intermediate flows that preferably sample closer objects than farther ones. Liu \textit{et al.}~\cite{22} proposed a cycle consistency neural network in which the synthesized frames are asserted to be more reliable if they could be used to reconstruct the input frames accurately. 
Park \textit{et al.}~\cite{Park} proposed a VFI method by considering the exceptional motion patterns. Lee \textit{et al.}~\cite{Lee} proposed a new warping module, namely Adaptive Collaboration of Flows (AdaCoF), to estimate both kernel weights and offset vectors for each target pixel to synthesize the missing frame.

\subsection{ST-SR for video}\label{subsec2.3}

The intuitive way to perform ST-SR is to combine the SSR and TSR alternatively. However, this approach treats each context, space and time independently. Another way is to simultaneously generate the output video with high resolution and high frame rate. Since pixels and frames are missed in low spatial and temporal resolution video, ST-SR for video is a highly ill-posed inverse problem. 

Shechtman \textit{et al.}~\cite{23} combined the sub-pixel and sub-frame misalignments information extracted from multiple videos with a directional space-time smoothness regularization to constrain the ill-posed problem. 
Mudenagudi \textit{et al.}~\cite{24} modeled the high-resolution video as a Markov random field and used a maximum posteriori estimation~\cite{25} as the final solution by using graph-cut~\cite{26}. 
Takeda \textit{et al.}~\cite{27} exploited both local spatial orientations and local motion vectors into account and adaptively constructed a suitable filter at every position of interest. Shahar \textit{et al.}~\cite{28} proposed an approach by combining information from multiple space-time patches to super resolve input videos. Although great achievement has been made by the existing methods, the performance still needs to be improved, especially for videos with complex spatial textures and temporal motions. 
Kang \textit{et al.}~\cite{b1} introduced a novel weighting scheme to effectively fuse all input frames without requiring explicit motion compensation. This approach improves the processing efficiency. 
Dutta \textit{et al.}~\cite{b2} utilized quadratic modeling to interpolate in LR space. They also reused the flowmaps and blending mask used to synthesizing the LR interpolated frame in HR space with bilinear upsampling. 
Xiang \textit{et al.}~\cite{b3} proposed a feature temporal interpolation network to capture local temporal contexts when interpolating LR frame features in missing LR video frames. They also introduced a deformable ConvLSTM to align and aggregate temporal information simultaneously, resulting in better leveraging of global temporal contexts. 
Haris \textit{et al.}~\cite{b4} developed a deep neural network that utilized direct lateral connections between multiple resolutions to present  rich multi-scale features during training. 
Xu \textit{et al.}~\cite{b5} introduced a temporal modulation block that modulates deformable convolution kernels for controllable feature interpolation. They also presented a locally temporal feature comparison module.

To improve the quality of ST-SR, we propose an end-to-end multi-branch-based neural network to directly learn the mapping between the input video with low spatial and temporal resolution and the output video with high spatial and temporal resolution.

%All papers must be written in UK English. If English is not
%your first language, you should ask an English-speaking
%colleague to proofread your paper. Papers that fail to meet
%basic standards of literacy are likely to be unsubmitted by
%the Editorial Office.

\section{Proposed method}\label{sec3}

Given a low-resolution and low frame rate video $\textbf{\textit{V}}^{in}= \left\{\textbf{\textit{I}}_{1}^{low }, \textbf{\textit{I}}_{3}^ {low },  \ldots, \textbf{\textit{I}}_{2 t+1}^{low }\right\}$ with $ N $ frames, the goal of ST-SR is to generate the corresponding high resolution and high frame rate video with $ 2N-1 $ frames, $\textbf{\textit{V}}^{out}= \left\{\textbf{\textit{I}}_{1}^{high}, \textbf{\textit{I}}_{2}^{high }, \textbf{\textit{I}}_{3}^{high },  \ldots, \textbf{\textit{I}}_{2 t+1}^{high }\right\}$, where $ \textbf{\textit{I}} $ denotes a frame, the superscripts ``\textit{low}'' and ``\textit{high}'' denote ``\textit{low}'' and ``\textit{high}'' resolutions, respectively, and the subscript ``\textit{t}'' is the time variable. 

The proposed Cuboid-Net, as shown in Fig.~\ref{Net}, contains four modules, i.e., MBFE module, MBR module, QE module, and CFQE module for interpolated frames only.
The MBFE module, comprising of three multi-feature blocks (MFBs), efficiently extracts features by observing the input video from various dimensions.
The MBR module is utilized for the reconstruction and fusion of high-resolution slices from their corresponding dimensions, comprising of three reconstruction blocks (RBs) and a fusion Block.
The first stage QE module enhances the output of the MBR module frame by frame, while the second stage CFQE module is used to further improve the quality of the high-resolution interpolated frames specifically with feature blocks (FBs) and convolution block attention module (CBAM)~\cite{29}.

\subsection{MBFE Module}\label{subsec3.1}

The MBFE module is the key of the proposed Cuboid-Net. It extracts features by effectively observing the input video from different dimensions. We take the input video as a cuboid with horizontal, vertical, and temporal dimensions. The horizontal and the vertical dimensions are $ H $ and $ W $, correspond to spatial resolution of the input video, the temporal dimension corresponds to the number of frames ($ N $) of the input video. To fully exploit the pixel correspondence in both spatial and temporal domains, we cut the cuboid horizontally, vertically, and temporally to generate three sets of sliced images, as shown in Fig.~\ref{slice}. Then, the features of each sliced image are extracted by using MFB independently. Given the input video $ \textbf{\textit{V}}^{in} $, $ \textbf{\textit{V}}_{1}$, $ \textbf{\textit{V}}_{2} $ and $ \textbf{\textit{V}}_{3} $ can be obtained after slicing operations in three directions:

\begin{figure*}[!h]
	\centering
	\includegraphics [scale=0.45]{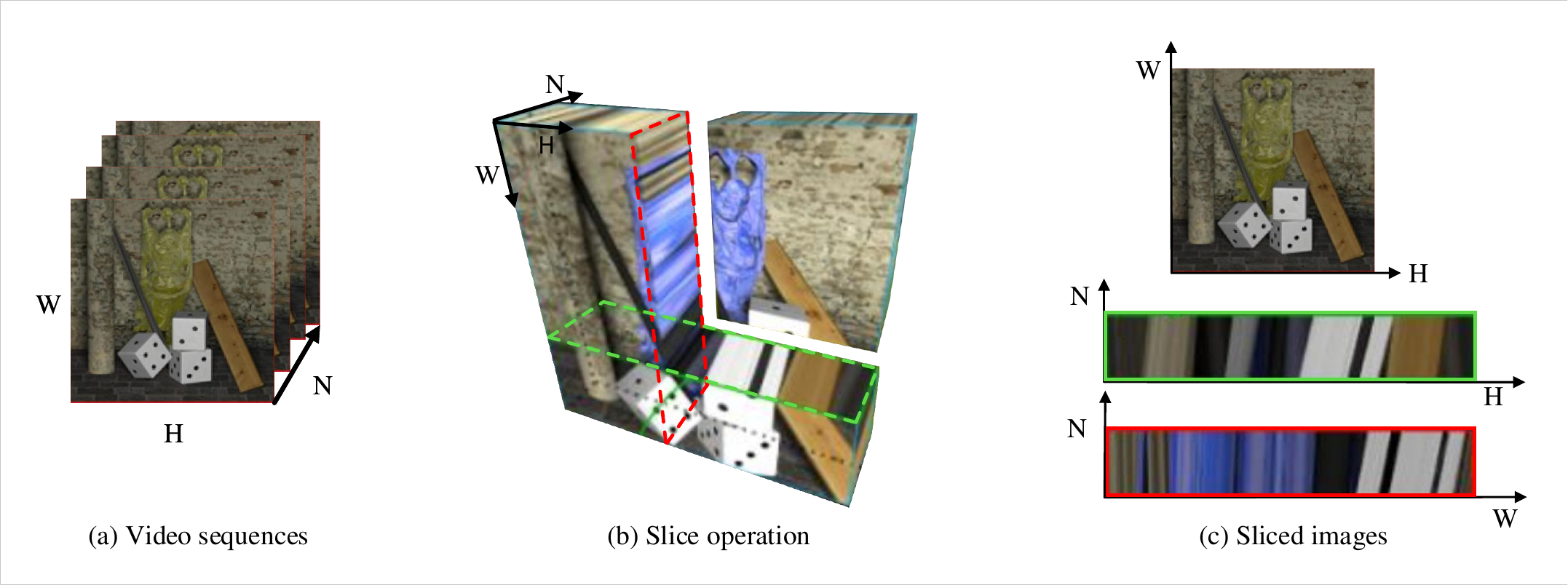}
	%\centerline {\includegraphics [scale=0.4]{fig3.pdf}}
	\caption{Slice illustration, (a) stack the input video to form a cuboid, (b) cut the cuboid into slices along with three directions, (c) obtained sliced images in the three sliced video.}
	\label{slice}
\end{figure*}

\begin{equation}
	\begin{aligned}
		\textbf{\textit{V}}^{in} 
		& = \textbf{\textit{V}}_{1} = \left \{ \textbf{\textit{I}}_{1}^{1},\cdots, \textbf{\textit{I}}_{i}^{1}, \cdots, \textbf{\textit{I}}_{N}^{1}  \right \} \\
		& = \textbf{\textit{V}}_{2} = \left \{ \textbf{\textit{I}}_{1}^{2},\cdots, \textbf{\textit{I}}_{j}^{2}, \cdots, \textbf{\textit{I}}_{W}^{2}  \right \} \\
		& = \textbf{\textit{V}}_{3} = \left \{ \textbf{\textit{I}}_{1}^{3},\cdots, \textbf{\textit{I}}_{k}^{3}, \cdots, \textbf{\textit{I}}_{H}^{3}  \right \},
	\end{aligned}
\end{equation}
where $ \textbf{\textit{I}}_{i}^{1}, i\le N $ is the sliced image with the resolution of $ (H, W) $, $ \textbf{\textit{I}}_{j}^{2}, j\le W $ is the sliced image with the resolution of $ (H, N) $ and $ \textbf{\textit{I}}_{k}^{3}, k\le H $ is the sliced image with the resolution of $ (W, N) $, as shown in Fig.~\ref{slice} (c).

\begin{figure}[!h]
	\centering
	\subfigure[]{
		\centering
		\includegraphics[width=\linewidth]{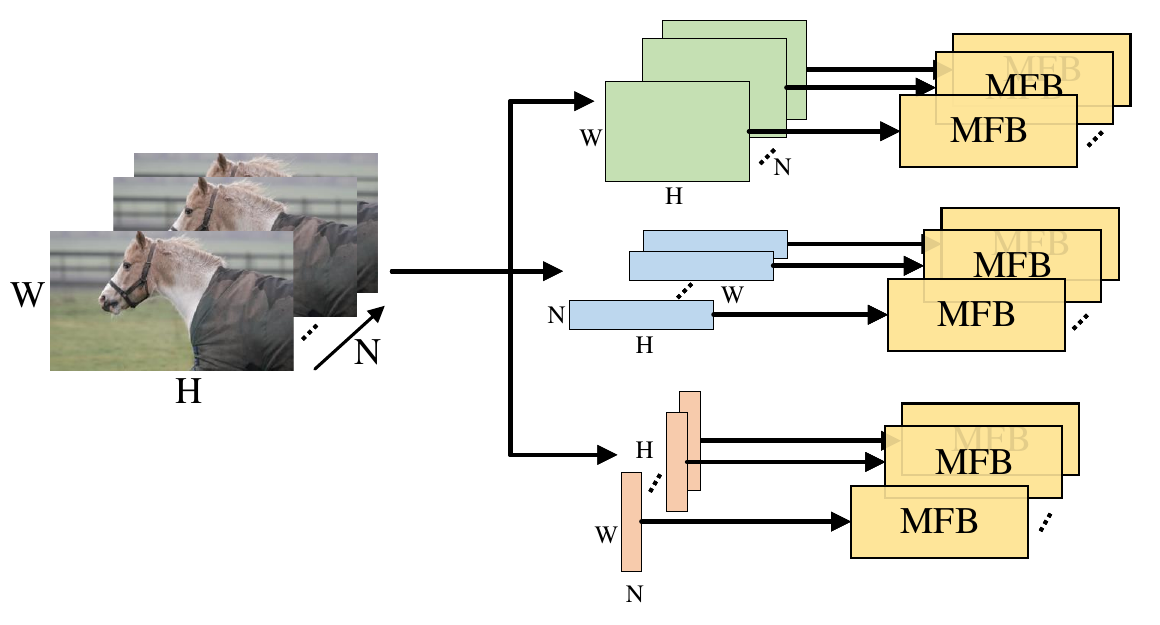}
		\label{4-a}
	}
	\subfigure[]{
		\centering
		\includegraphics[width=\linewidth]{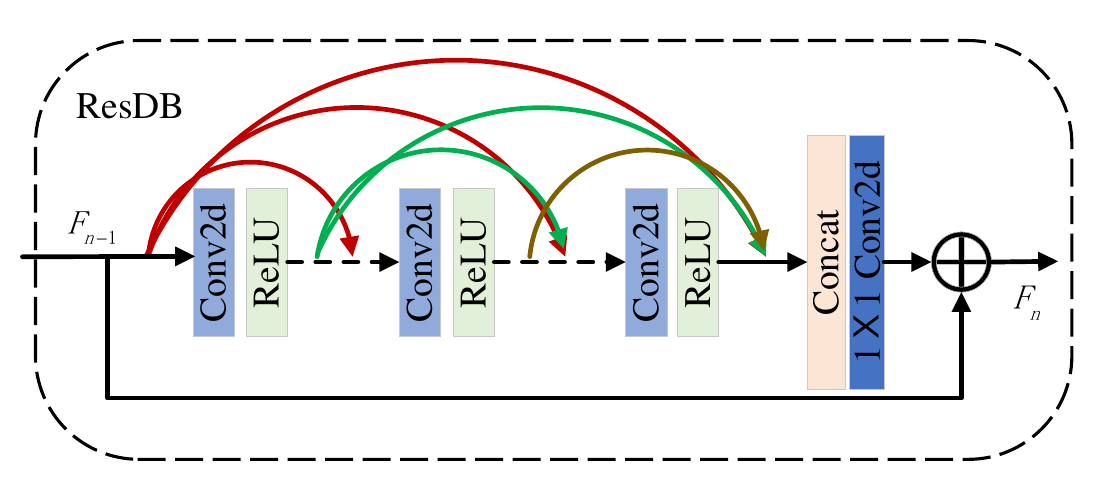}
		\label{4-b}
	}
	\subfigure[]{
		\centering
		\includegraphics[width=\linewidth]{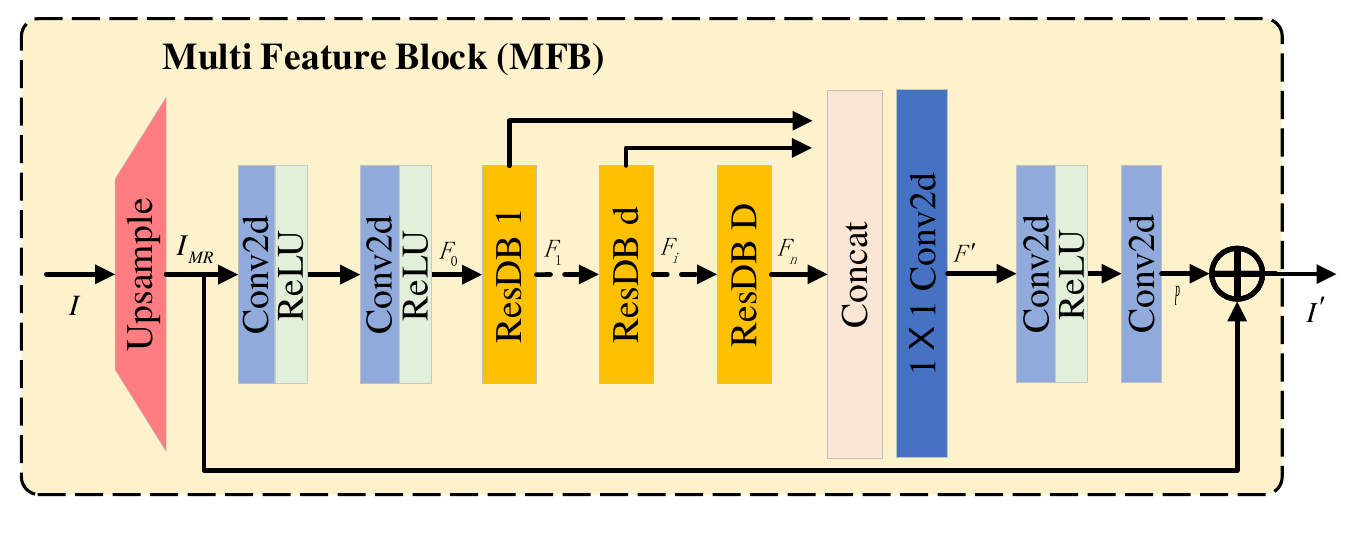}
		\label{4-c}
	}
	\centering
	\caption{Structure of MBFE module, (a) the overall structure, (b) the structure of ResDB block, (c) the structure of multi-feature block (MFB).}
	\label{MBFE}
\end{figure}

The detailed structure of the MBFE module is shown in Fig.~\ref{MBFE}. This module is used to extract hybrid features for spatial and temporal simultaneously. In Fig.~\ref{4-a}, the MBFE module consists of three branches of slice and MFB blocks. Each branch is for a sliced image set $ \textbf{\textit{V}}_{m},\, m=1, 2, or\,3, $ generated by cutting the input video cuboid. The three branches adopt the same network structure as shown in Fig.~\ref{4-c}. The MFB block first improves the resolution of each sliced image $ \textbf{\textit{I}} $ by using a simple bicubic interpolation filter. The initial up-sampled image $ \textbf{\textit{I}}_{MR} $ is then used as the basic information for the following residual learning in which two 2D convolution layers for shallow feature extraction and a series of residual dense block (ResDB) for deep feature extraction are used. The structure of ResDB is shown in Fig.~\ref{4-b}. It consists of three convolution layers followed by a ReLU with dense connection and a 1$ \times $1 convolution layer. 
In the ResDB, the skip connections are used between blocks which add the output of the previous block and current block to the next block. The output features of all the ResDB blocks are concatenated so that the long-term features can also be reserved. A 1$ \times $1 convolutional layer with Leaky ReLU is then used to reduce the channel of the extracted features. After that, 2D convolution layers are used to predict the residual information of the input image. 

The ResDB in Fig.~\ref{4-b} can be formulated as
\begin{equation}
	\textbf{\textit{F}}_{n} = \mathrm{ResDB}(\textbf{\textit{F}}_{n-1}) = \textbf{\textit{F}}_{n-1} + \mathrm{R}(\mathrm{D}(\textbf{\textit{F}}_{n-1})),
\end{equation}
where $ \textbf{\textit{F}}_{n} $ is the output, $ \textbf{\textit{F}}_{n-1} $ is the input, $ \mathrm{D}(\cdot) $ is the dense connection operation, and $ \mathrm{R}(\cdot) $ is the concatenation operation followed by a 1$ \times $1 convolution for channel reduction.

For the MFB block in Fig.~\ref{4-c}, the input is $ \textbf{\textit{I}} $ and the output is $ {\textbf{\textit{I}}}' $. $ \textbf{\textit{I}}_{MR} $ is the initial up-sampled image by using a simple bicubic interpolation filter. $ \textbf{\textit{F}}_{0} $ is the extracted shallow feature extraction, and $ \textbf{\textit{F}}_{i} $ is the extracted deep feature by ResDB, $ \textbf{\textit{P}} $ is the residual information. The MFB block can be formed as follows,

\begin{equation}
	\textbf{\textit{F}}_{0} = \mathrm{CR}[\mathrm{CR}(\textbf{\textit{\textbf{\textit{I}}}}_{MR})], 
\end{equation}
\begin{equation}
	\textbf{\textit{F}}_{n} = \mathrm{ResDB}(\textbf{\textit{F}}_{n-1}), 
\end{equation}
\begin{equation}
	{\textbf{\textit{F}}}' = \mathrm{R}(\left \{ \textbf{\textit{F}}_{1}, \cdots, \textbf{\textit{F}}_{n} \right \}), 
\end{equation}
\begin{equation}
	\textbf{\textit{P}} = \mathrm{C}[\mathrm{CR}({\textbf{\textit{F}}}')],
\end{equation}
\begin{equation}
	{\textbf{\textit{I}}}' = \textbf{\textit{I}}_{MR} +\textbf{\textit{P}},
\end{equation}
where $ \mathrm{CR}(\cdot) $ is the 2D convolution followed by a ReLU, $ \mathrm{R}(\cdot) $ is the concatenation operation followed by a 1$ \times $1 convolution to reduce the channel, and $ \mathrm{C}(\cdot) $ is the 2D convolution operation.

Finally, the output of each branches of the MBEF module can be written as
\begin{equation}
	\begin{aligned}
		{\textbf{\textit{V}}_{m}}' &=  \mathrm{MBFE}(\textbf{\textit{V}}^{in}) \\
		&= \left \{ {\textbf{\textit{I}}_{1}^{m}}', {\textbf{\textit{I}}_{2}^{m}}',{\textbf{\textit{I}}_{3}^{m}}'\cdots \right \},  	\quad m=1,2,or\,3,
	\end{aligned}
\end{equation}
where the $ m $ denotes a branch.

\subsection{MBR Module}\label{subsec3.2}

\begin{figure}[!h]
	\centering
	\includegraphics [scale=0.45]{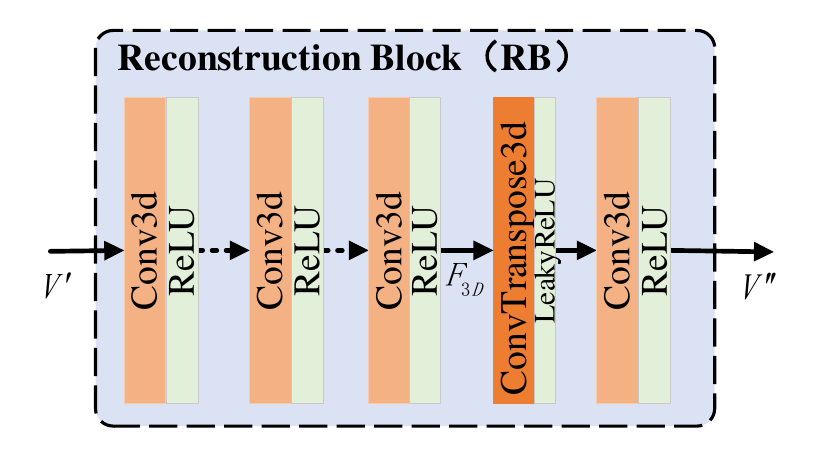}
	%%\centerline {\includegraphics [scale=0.5]{fig5.pdf}}
	\caption{The structure of RB module.}
	\label{RB}
\end{figure}

The MBR module is composed of three RBs and a fusion block which is used to reconstruct and fuse high resolution video from the three branches, as shown in Fig.~\ref{Net}. 
The extracted features of each branch are used as the input of the RB. Specifically, the RB is made up of a series of 3DConv layers, a single 3D transposed convolutional layer followed by a Leaky ReLU layer, and a 3DConv, as shown in Fig.~\ref{RB}. The output features of the three RBs are fused together to generate the initial reconstruction of the super-resolved video. Since the output features of the three RBs contain both temporal and spatial information, 3D convolution is used to jointly fuse them. The RB can be formed as follows,

%\vspace{-0.5cm}

\begin{equation}
	\textbf{\textit{F}}_{3D} = \mathrm{C}_{3D}\cdot \mathrm{C}_{3D}\cdots \mathrm{C}_{3D}({\textbf{\textit{V}}_{m}}'),
\end{equation}

\begin{equation}
	{\textbf{\textit{V}}_{m}}'' = \mathrm{C}_{3D}[\mathrm{T}_{3D}(\textbf{\textit{F}}_{3D})],
\end{equation}
where $ \textbf{\textit{F}}_{3D} $ is the features extracted by a series of 3DConv layers, $ \mathrm{C}_{3D}(\cdot) $ is the 3D convolution operation followed by a ReLU, and $ \mathrm{T}_{3D}(\cdot) $ is the ConvTranspace3d operation followed by a LeakyReLU. The output of the MBR module can be written as

\begin{equation}
	\textbf{\textit{Out}} = \mathrm{MBR}(\left \{ {\textbf{\textit{V}}_{1}}', {\textbf{\textit{V}}_{2}}',{\textbf{\textit{V}}_{3}}'\right \}  ).
\end{equation}

\subsection{First Stage Quality Enhancement (QE) Module}\label{subsec3.3}

\begin{figure}[!h]
	\centering
	\includegraphics [scale = 0.45]{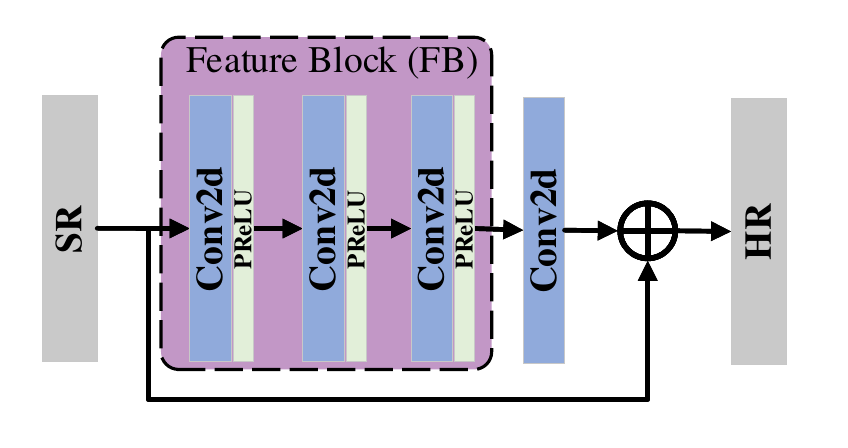}
	%%\centerline {\includegraphics [scale=0.5]{fig5.pdf}}
	\caption{The structure of QE module.}
	\label{QE}
\end{figure}

We then perform the first stage quality improvement for the output of the MBR module frame by frame. Inspired by the SRCNN~\cite{SRCNN}, the structure of the QE module consists of three successive convolutional layers with Parametric Rectified Linear Unit (PReLU) to maintain more detailed features, a single convolutional layer, and a skip connection for residual learning, as shown in Fig.~\ref{QE}. 
\vspace{-0.5cm}

\subsection{Second Stage Cross Frame Quality Enhancement (CFQE) module}\label{subsec3.4}

\begin{figure}[!h]
	\centering
	\includegraphics [width=\linewidth]{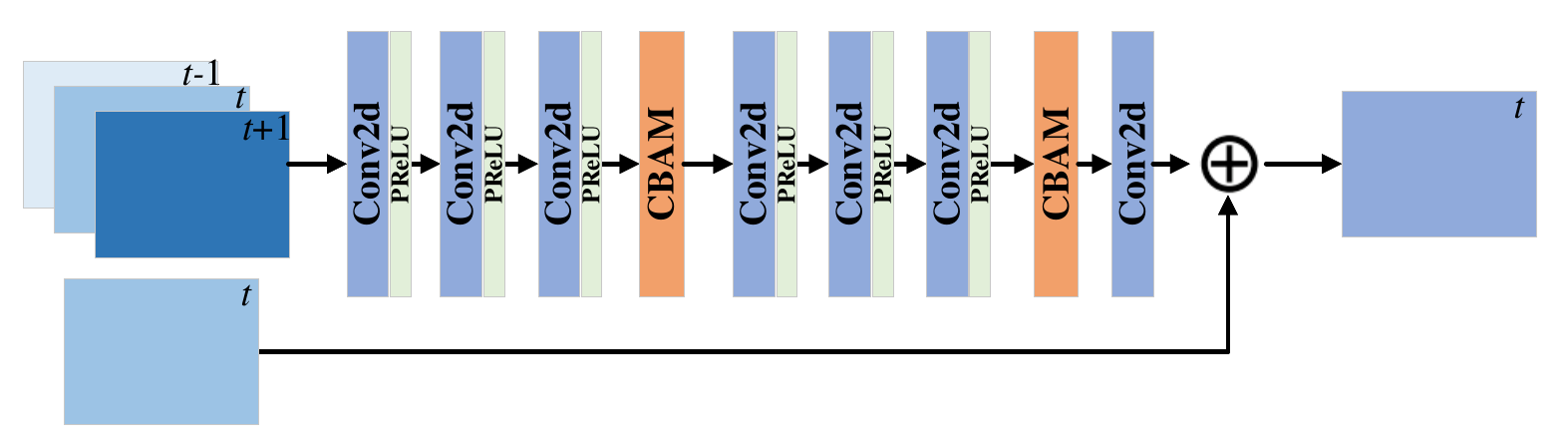}
	%%\centerline {\includegraphics [scale=0.4]{fig6.pdf}}
	\caption{The structure of CFQE module.}
	\label{CFQE}
\end{figure}

\begin{figure}[!h]
	\centering
	\includegraphics [scale=0.33]{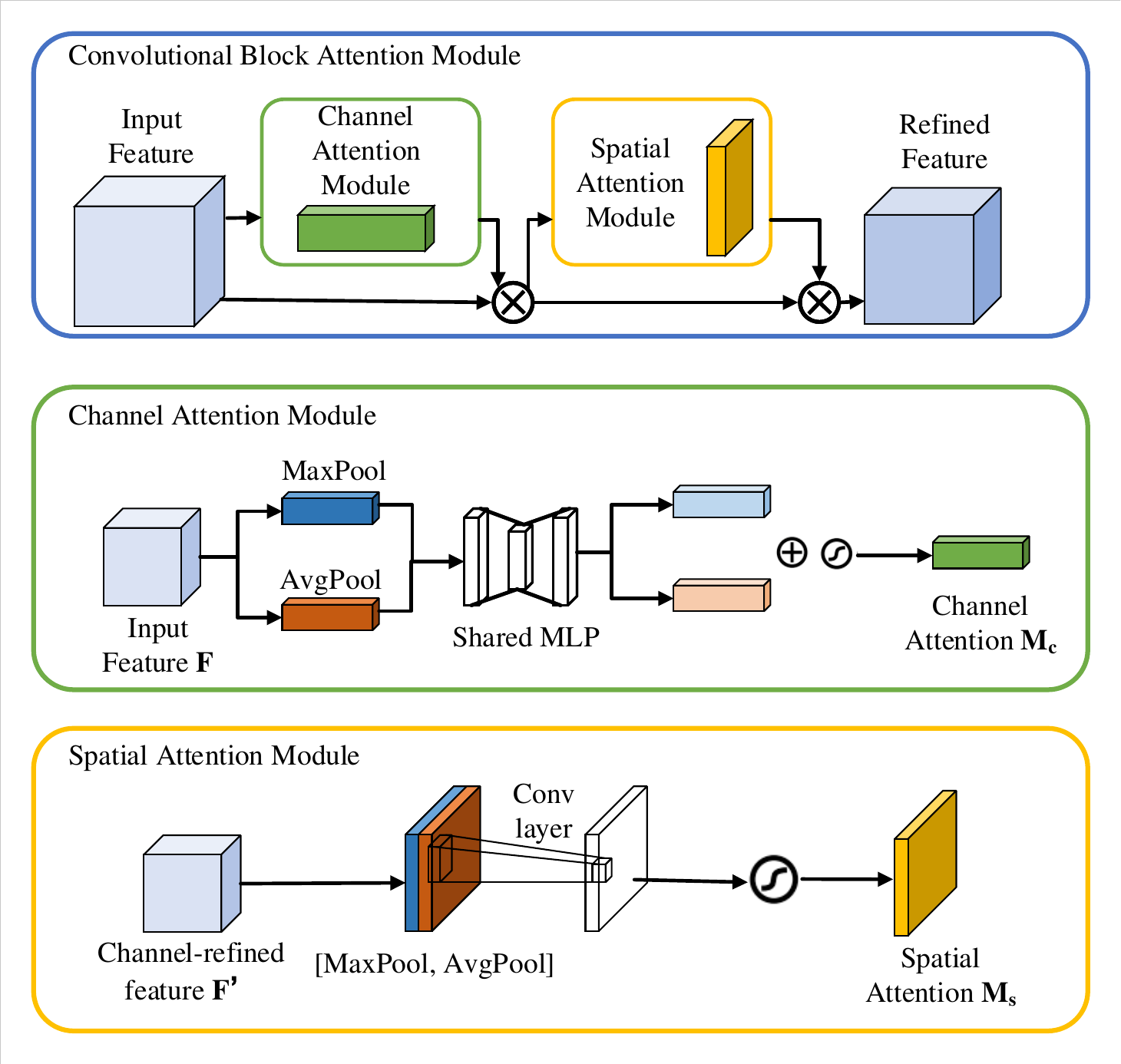}
	%%\centerline {\includegraphics [scale=0.3]{fig7.pdf}}
	\caption{The structure of Convolutional Block Attention Module (CBAM).}
	\label{CBAM}
\end{figure}

Due to the complex motion in videos, the quality of the temporally interpolated frames are usually lower than those of the spatially super-resolved frames, which affects the perceptual quality greatly. To address this problem, we design a CFQE module to further improve the image quality of the interpolated frames, as shown in Fig.~\ref{CFQE}. The core of the CFQE module is to learn the residual information of the temporally interpolated frames (time \textit{t} in Fig.~\ref{CFQE}) by exploring the features of its neighboring frames (time \textit{t}+1 and \textit{t}-1 in Fig.~\ref{CFQE}). The CFQE consists of seven convolution layers and two CBAMs~\cite{29} (see Fig.~\ref{CBAM}). The first six layers associated with a ReLU activation function are used for feature extraction, whereas the last layer is used for residual image reconstruction. 

The CBAM module is designed to address two crucial problems: spatial and channel attention, as illustrated in Fig.~\ref{CBAM}. It enriches feature representations by dynamically recalibrating feature maps, thereby capturing both local and global dependencies. Within the CBAM module, the spatial attention mechanism effectively emphasizes pertinent spatial regions within individual feature maps. This is achieved by considering the interactions among various spatial locations and accentuating the most informative regions. On the other hand, the channel attention mechanism assesses the importance of different feature channels. It assigns varying degrees of significance to these channels based on their contributions to the overall feature representation. By integrating spatial and channel attention, the CBAM module refines feature maps by accentuating relevant spatial regions and assigning prominence to essential feature channels.

Given an intermediate feature map, the CBAM module sequentially infers attention maps along two separate dimensions (channel and spatial), which are then multiplied with the input feature map to fully explore the temporal correspondence features between successive frames.

\section{Experimental Results and Ablation Study}\label{sec4}

\subsection{Data sets and training configuration}\label{subsec4.1}

To verify the performance of the proposed method, we used Vimeo-90K~\cite{90K} as the training set which includes more than 600000 videos with size of 7 (number of frames)$ \times $ 448 (spatial width) $ \times $ 256 (spatial height). The Vid4~\cite{Vid4} and Vimeo-90K~\cite{90K} datasets were used for test. Similar to~\cite{34}, to measure the performance under different motion conditions, we split the Vimeo 90K dataset into fast motion (1255 videos) group, medium motion (4977 videos) group, and slow motion (1613 videos) group, respectively. To train the proposed network, we first generated low-resolution videos by spatially and temporally sub-sampling. Specifically, the spatial resolution was down-sampled to 112$ \times $64 through bi-cubic interpolation method, while the even frames were deleted directly to reduce the temporal resolution. 
We used the widely used Peak Signal-to-Noise Ratio (PSNR) and Structural Similarity Index (SSIM)~\cite{35} to assess the quality of the generated high-resolution videos. The PSNR quantifies the mean squared error (MSE) between a reference image and a distorted version as,
\begin{equation}
	PSNR=10 \cdot \log _{10}\left(\frac{M A X^2}{M S E}\right),
\end{equation}	
where $ MAX $ represents the maximum possible pixel value (i.e. 255 for a pixel with 8 bits) and $ MSE $ denotes the Mean Squared Error, the average of the squared pixel differences between the ground truth image and the reconstructed image.
The higher the PSNR value, the closer the distorted image is to the original.
SSIM measures the similarity between two images, considering structural information, luminance, and contrast. SSIM takes into account human visual perception and provides a better reflection of perceived image quality. The formula for SSIM is given as follows~\cite{35} :
\begin{equation}
	\operatorname{SSIM}(x, y)=\frac{\left(2 \mu_x \mu_y+C_1\right)\left(2  \sigma_{x y}+C_2\right)}{\left(\mu_x^2+\mu_y^2+C_1\right) \left(\sigma_x^2+\sigma_y^2+C_2\right)},
\end{equation}
where $ x $ and $ y $ are the compared images, $ \mu_x $ and $ \mu_y $ are the mean values of $ x $ and $ y $, $ \sigma_x $ and $ \sigma_y $ are the standard deviations of $ x $ and $ y $, $ \sigma_{x y} $ is the covariance of $ x $ and $ y $, $ C_1 $ and $ C_2 $ are constants to stabilize the division in the formula.
SSIM yields a value within the range of 0 to 1, where 1 signifies perfect perceptual quality relative to the ground truth.

The proposed model was implemented on PyTorch platform and trained by a graphical card with 2080Ti GPU. In the implementation, we randomly cropped the videos for training to patches with size of 32 $ \times $ 32$ \times $4 as input and used the corresponding ground truth video patch with size of (32$ \times $4)$ \times $ (32 $ \times $ 4)$ \times $7 as the labels. During training, the batch size was set to 8, the Adam optimizer with $ \beta _1=0.5 $ and $ \beta _2=0.99 $~\cite{36} was adopted, the learning rate was initialized to 0.0001 and decreased by a factor of 0.5 after each 60 epochs. The L2 loss was adopted as the loss function.

We evaluated the performance in three aspects, i.e., SSR quality, TSR quality, and ST-SR quality. The SSR quality is for the spatially reconstructed frames only, the TSR quality is for the temporally reconstructed frames only, while the ST-SR quality is for all the reconstructed frames.

\subsection{Ablation Study}\label{subsec4.2}

\subsubsection{Effect of the number of ResDBs}\label{subsubsec4.2.1}

The proposed ResDBs play an important role in the MBFE module. To verify the performance of ResDB, we studied the effect of the number of ResDBs, as shown in Tables~\ref{numResDBs} and~\ref{numResDBs-4}. In general, the quality can be improved by more ResDBs. For the test set Vimeo90K, when the number of ResDBs was increased from 3 to 9, the PSNR improvements were 1.77 dB for ST-SR, 2.12 dB for SSR, and 1.43 dB for TSR. When the number of ResDBs was raised from 7 to 9, the PSNR improvements were 0.43 dB for ST-SR, 0.44 dB for SSR, and 0.41 dB for TSR. For the Vid4 test set, when the number of RABs was raised from 7 to 9, the PSNR improvements were only 0.08 dB for ST-SR, 0.08 dB for SSR, and 0.09 dB for TSR. To balance the computational complexity and performance, we set the number of ResDBs to 7 for Vid4 and 9 for Vimeo90K.

\begin{table}[!h]
	\centering
	\caption{Performance of the number of ResDBs on Vimeo90K.}
	\renewcommand\arraystretch{1.5}
	%\resizebox{\linewidth}{!}{ 
		\begin{tabular}{c|c|c|c|c|c|c}
			\hline
			\multirow{2}{*}{ResDBs} & \multicolumn{2}{c|}{  ST-SR }  & \multicolumn{2}{c|}{ SSR }& \multicolumn{2}{c}{ TSR }\\
			\hhline{~------}
			& PSNR & SSIM & PSNR & SSIM & PSNR & SSIM \\
			\hline
			3       & 29.31 & 0.876 & 30.02 & 0.900 & 28.59 & 0.824\\
			5       & 30.01 & 0.899 & 30.96 & 0.922 & 29.03 & 0.867   \\
			7       & 30.65 & 0.918 & 31.70 & 0.934 & 29.61 & 0.875  \\
			9  & \textbf{31.08} & \textbf{0.931} & \textbf{32.14} & \textbf{0.941} & \textbf{30.02} & \textbf{0.903} \\
			\hline
		\end{tabular}
		%}
	\label{numResDBs}
\end{table}

\begin{table}[!h]
	\centering
	\caption{Performance of the number of ResDBs on Vid4.}
	\renewcommand\arraystretch{1.5}  
	%\resizebox{\linewidth}{!}{
		\begin{tabular}{c|c|c|c|c|c|c}
			\hline
			\multirow{2}{*}{ResDBs} & \multicolumn{2}{c|}{ ST-SR }  & \multicolumn{2}{c|}{ SSR }& \multicolumn{2}{c}{ TSR }\\
			\hhline{~------}
			& PSNR & SSIM & PSNR & SSIM & PSNR & SSIM \\
			\hline
			3       & 28.54 & 0.862  & 29.08  & 0.867 & 27.48 & 0.825\\
			5       & 29.23 & 0.875  & 30.02  & 0.892 & 27.82 & 0.831   \\
			7       & 29.61 & 0.879  & 30.67 & 0.905 & 28.37 & 0.849   \\
			9      & \textbf{29.69} & \textbf{0.882} & \textbf{30.75}  &\textbf{0.910} & \textbf{28.46} & \textbf{0.852} \\
			\hline
		\end{tabular}
		%	}
	\label{numResDBs-4}
\end{table}

\subsubsection{Effect of the number of Conv3d layers}\label{subsubsec4.2.2}

To verify the performance of Conv3d in the MBR module, we studied the effect of the number of Conv3d layers, as shown in Table~\ref{num3D}. The Conv3d layers in RB are mainly used to extract space-time fusion features. We can see that the quality can be improved by more Conv3d layers. For the Vimeo90K test set, when the number of Conv3d layers was increased from 1 to 5, the PSNR improvements were 0.33 dB, 0.35 dB, and 0.3 dB for ST-SR, SSR, and TSR, respectively.

\begin{table}[!h]
	\centering
	\caption{Performance of the number of Conv3d layers.}
	\renewcommand\arraystretch{1.5}  
	\resizebox{\linewidth}{!}{
		\begin{tabular}{c|c|c|c|c|c|c}
			\hline
			\multirow{2}{*}{Conv3d layers} & \multicolumn{2}{c|}{  ST-SR }  & \multicolumn{2}{c|}{ SSR }& \multicolumn{2}{c}{ TSR }\\
			\hhline{~------}
			& PSNR & SSIM & PSNR & SSIM & PSNR & SSIM \\
			\hline
			1      & 30.75 & 0.920  & 31.79  & 0.934 & 29.72 & 0.881\\
			3      & 30.94 & 0.922  & 31.96  & 0.937 & 29.84 &  0.887  \\
			5      &\textbf{31.08} & \textbf{0.931} & \textbf{32.14} & \textbf{0.941} & \textbf{30.02} & \textbf{0.903}\\
			\hline
		\end{tabular}
	}
	\label{num3D}
\end{table}

\subsubsection{Effect of the network modules}\label{subsubsec4.2.3}

To evaluate the efficiency of each module, we trained and tested the network by using the corresponding modules: MBFE+MBR, MBFE+MBR+QE, MBFE+MBR+QE+CFQE. Experimental results were shown in Table~\ref{tab1} for Vid4 dataset and Table~\ref{tab2},~\ref{network} for Vimeo 90K, respectively. 

As shown in Tables~\ref{tab1} and~\ref{tab2}, for the Vid4 and Vimeo 90K datasets, we can see that there are 0.17 dB and 0.11 dB PSNR improvements for ST-SR, 0.23 dB and 0.15 dB for SSR, and 0.08 dB and 0.07 dB for TSR, respectively, when the QE module is added after the MBFE and MBR module, illustrating the effectiveness of the QE module. The CFQE module is designed to improve the quality of the interpolated frames. From the results in the two tables, we can see that CFQE is also effective, and the TSR quality can be improved by 0.04 dB and 0.07 dB on Vid4 and Vimeo 90K datasets, respectively.
Table~\ref{network} compared the quality of fast motion group, medium motion group, and slow motion group based on Vimeo 90K dataset, respectively. From all the tables, we can see that the performance of MBFE+MBR+QE+CFQE is the best in terms of SSR, TSR and ST-SR quality.

\begin{table*}[!h]
	\centering
	\caption{Comparison result on Vid4.}
	\label{tab1}
	\renewcommand\arraystretch{1.5}  
	%\resizebox{\linewidth}{!}{
		\begin{tabular}{c|c|c|c|c|c|c|c|c}
			\hline
			\multirow{2}{*}{Methods} & \multicolumn{2}{c|}{ ST-SR }  & \multicolumn{2}{c|}{ SSR }& \multicolumn{2}{c|}{ TSR } & \multirow{2}{*}{Parameters} & \multirow{2}{*}{Runtime}\\
			\hhline{~------}
			& PSNR & SSIM & PSNR & SSIM & PSNR & SSIM & & \\
			\hline
			MBFE+MBR          & 29.41 & 0.856 & 30.39 & 0.901 & 28.25 & 0.834 & 17.7& 12.79\\
			MBFE+MBR+QE       & 29.58 & 0.881 & 30.62 & 0.908 & 28.33 & 0.841 & 17.8 &  12.85\\
			MBFE+MBR+QE+CFQE   & \textbf{29.69} & \textbf{0.882} & \textbf{30.75}  &\textbf{0.910} & \textbf{28.46} & \textbf{0.852}& 18.1 & 12.96\\
			\hline
		\end{tabular}
%	}
\end{table*}

\begin{table*}[!h]
	\centering
	\caption{Comparison result on Vimeo90K}
	\label{tab2}
	\renewcommand\arraystretch{1.5}  
	%\resizebox{\linewidth}{!}{
		\begin{tabular}{c|c|c|c|c|c|c|c|c}
			\hline
			\multirow{2}{*}{Methods} & \multicolumn{2}{c|}{  ST-SR }  & \multicolumn{2}{c|}{ SSR }& \multicolumn{2}{c|}{ TSR }& \multirow{2}{*}{Parameters} & \multirow{2}{*}{Runtime}\\
			\hhline{~------}
			& PSNR & SSIM & PSNR & SSIM & PSNR & SSIM & & \\
			\hline
			MBFE+MBR          &30.80 & 0.8975 & 32.25 & 0.9215 & 28.85 & 0.8657 & 17.7& 13.38\\
			MBFE+MBR+QE       & 30.91 & 0.8986 & 32.40 & 0.9222 & 28.92 & 0.8670 & 17.8 &13.45\\
			MBFE+MBR+QE+CFQE   &\textbf{31.08} & \textbf{0.931} & \textbf{32.14} & \textbf{0.941} & \textbf{30.02} & \textbf{0.903} & 18.1 & 13.58 \\
			\hline
		\end{tabular}
%	}
\end{table*}

\begin{table*}[!h]
	\centering
	\caption{Comparison results of modules on Vimeo90K}
	\label{network}
	\renewcommand\arraystretch{1.5}  
	%\resizebox{\linewidth}{!}{
		\begin{tabular}{c|c|c|c|c|c|c|c}
			\hline
			\multicolumn{2}{c|}{\multirow{2}{*}{Methods}} & \multicolumn{2}{c|}{ Vimeo-Fast }  & \multicolumn{2}{c|}{ Vimeo-Medium }& \multicolumn{2}{c}{ Vimeo-Slow }\\
			\hhline{~~------}
			\multicolumn{2}{c|}{} & PSNR & SSIM & PSNR & SSIM & PSNR & SSIM \\
			\hline
			MBFE+MBR   & \multirow{3}{*}{ST-SR}  & 31.31 & 0.8710 & 31.00 & 0.9048 & 29.76 & 0.8953 \\
			MBFE+MBR+QE  &     & 31.49 & 0.8729 & 31.11 & 0.9055 & 29.84 & 0.8966\\
			MBFE+MBR+QE+CFQE & & \textbf{31.50} & \textbf{0.8717} & \textbf{31.15} & \textbf{0.9055} & \textbf{29.89} & \textbf{0.8673}\\			
			\hline
			MBFE+MBR   &  \multirow{3}{*}{SSR} & 34.75 & 0.9374 & 32.30 & 0.9237 & 30.24 & 0.9025 \\
			MBFE+MBR+QE  &     & 35.02 & 0.9388 & 32.44 & 0.9243 & 30.31 & 0.9029\\
			MBFE+MBR+QE+CFQE & & \textbf{35.03} & \textbf{0.9390} & \textbf{32.47} & \textbf{0.9244} & \textbf{30.33} & \textbf{0.9031}\\
			\hline
			MBFE+MBR   & \multirow{3}{*}{TSR}  & 26.73 & 0.7826 & 29.28 & 0.8796 & 29.13 & 0.8857 \\
			MBFE+MBR+QE  &     & 26.68 & 0.7818 & 29.28 & 0.8789 & 29.19 & 0.8874\\
			MBFE+MBR+QE+CFQE & & \textbf{26.79} & \textbf{0.7849} & \textbf{29.35} & \textbf{0.8804} & \textbf{29.21} & \textbf{0.8882}\\
			\hline
		\end{tabular}
	%}
\end{table*}

\subsection{Compared with state-of-the-art Methods}\label{subsubsec4.3}

\begin{table*}[!t]
	\centering
	\caption{ST-SR quality.}
	\renewcommand\arraystretch{1.5}  
	%\resizebox{\linewidth}{!}{
		\begin{tabular}{l|c|c||l|c|c}
			\hline
			\makecell{Method \\ (SSR+TSR/ST-SR)} & PSNR↑ & SSIM↑ & 
			\makecell{Method \\(TSR+SSR/ST-SR)} & PSNR↑ & SSIM↑  \\
			\hline
			DBPN+ToFlow & 29.87 & 0.905 & ToFlow+DBPN & 28.82 & 0.887 \\
			TDAN+ToFlow & 30.41 & 0.918 & ToFlow+TDAN & 29.65 & 0.908 \\
			BasicVSR+ToFlow & 30.58 & 0.924 & ToFlow+BasicVSR & 30.22 & 0.915 \\
        	LatticeNet+ToFlow & 30.84 & 0.928 & ToFlow+LatticeNet & 30.62 & 0.924 \\
        	SwinIR+ToFlow & 30.91 & 0.929 & ToFlow+SwinIR & 30.73 & 0.926 \\
			RBPN+DAIN & 30.25 & 0.916 & DAIN+RBPN & 29.42 & 0.906 \\
			BasicVSR+DAIN & 30.72 & 0.925 & DAIN+BasicVSR & 30.45 & 0.919 \\
			LatticeNet+CAIN & 30.95 & 0.929 & CAIN+LatticeNet & 30.82 & 0.922 \\
			SwinIR+EDSC & 30.98 & 0.930 & EDSC+SwinIR & 30.84 & 0.928 \\
			\hline
			STVUN &   29.68  &  0.908 & STARnet &   30.81    &  0.924 \\
			TMNet &   30.92  &   0.928  & Ours  &  \textbf{31.08}     & \textbf{0.931} \\
			\hline
		\end{tabular}%
%	}
	\label{ST-SR}%
\end{table*}%

To compare the ST-SR quality, we alternatively combined the state-of-the-art SSR and TSR methods to obtain the ST-SR videos as anchors. Specifically, DBPN~\cite{37}, RBPN~\cite{38}, TDAN~\cite{5}, BasicVSR~\cite{44}, LatticeNet~\cite{LatticeNet} and SwinIR~\cite{SwinIR} were used for SSR, while ToFlow~\cite{34}, DAIN~\cite{21}, CAIN~\cite{CAIN} and EDSC~\cite{EDSC} were adopted for TSR. The compared ST-SR methods are STVUN~\cite{b1}, STARnet~\cite{b4}, and TMNet~\cite{b5}.
From Table~\ref{ST-SR}, we can see that the combination of SSR+TSR performs better than TSR+SSR. We can also see that both PSNR (31.08dB on average) and SSIM (0.931) of the proposed method are the best.

To illustrate the SSR performance of the proposed method, we compared the proposed method with four state-of-the-art methods, i.e., DBPN~\cite{37}, RBPN~\cite{38}, TDAN~\cite{5} and BasicVSR~\cite{44}, as shown in Table~\ref{tab6}. We can see that our method is only a little smaller than the state-of-the-art BasicVSR. 
Because the input of our method is videos with low resolution and low frame rate, whereas the input of the compared SSR methods are videos with low resolution and high frame rate.

\begin{table*}[!h]
	\centering
	\caption{SSR quality comparison}
	\label{tab6}
	\renewcommand\arraystretch{1.5}  
%	\resizebox{\linewidth}{!}{
		\begin{tabular}{c|c|c|c|c|c|c}
			%\begin{tabular}{c|c|c|c|c|c|c|c}
			\hline
			Methods &  Bicubic & DBPN &  RBPN & TDAN & BasicVSR & Ours \\
			\hline
			PSNR      & 27.89 & 29.83 & 30.97 & 32.42 & 33.02 & 32.81\\
			SSIM      & 0.887 & 0.913 & 0.938 & 0.943 & 0.957 & 0.952\\
			\hline
		\end{tabular}
			%}
\end{table*}

\begin{table*}[!h]
	\centering
	\caption{Computational complexity comparsion}
	\renewcommand\arraystretch{2.3}  
	%\resizebox{\linewidth}{!}{
		\begin{tabular}{c|c|c|c|c|c|c|c}
			\hline
			Methods & \makecell{ToFlow+ \\ DBPN} & \makecell{DBPN+\\DAIN}  & \makecell{RBPN+\\DAIN} & \makecell{TDAN+\\DAIN} & STARnet  & TMNet & Ours \\
			\hline
			\makecell{Parameters\\(million)}   & 20.2  & 38.4 & 36.7 & 26.2 &111.61 & \textbf{12.26} &18.1 \\
			\hline
			Runtime(s)      & 3.44  & 3.26 & 4.26 & 3.52 & 14.08 & 14.69 & \textbf{13.58}\\
			\hline
	\end{tabular}
%}
	\label{time}
\end{table*}

\begin{figure*}
	\centering
	\includegraphics [width=\linewidth]{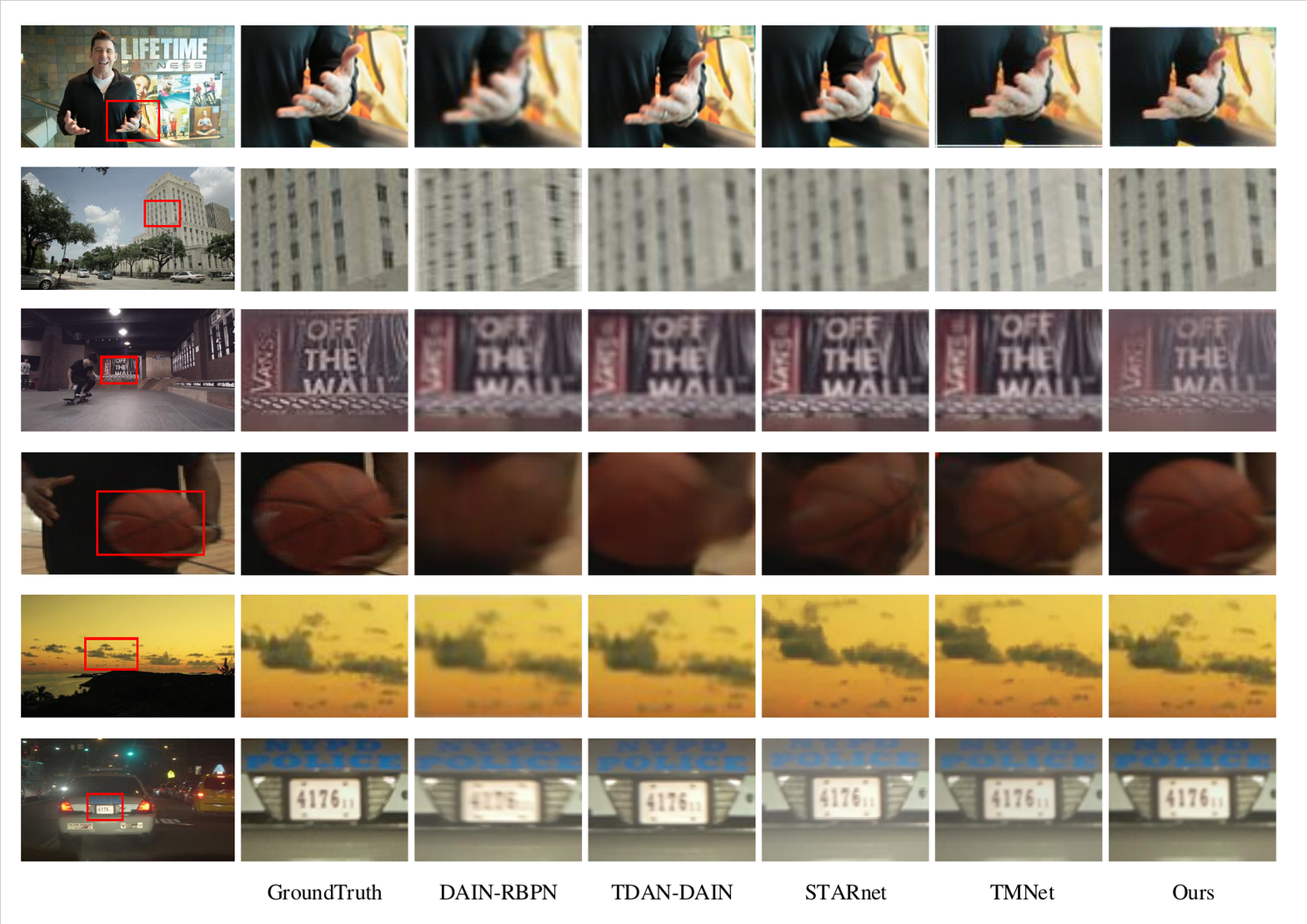}
	\caption{Visual results on ST-SR. Red arrows here and in the other figures indicates the highlighted area.}
	\label{Visual}
\end{figure*}

Moreover, we also investigate model sizes and runtime of different networks in Table~\ref{time}. For synthesizing high quality frames, SSR and TSR networks usually need very large frame reconstruction modules. Thus, the composed two-stage ST-SR networks will contain a huge number of parameters, while our one-stage model needs fewer parameters. From Table~\ref{time}, we can see that the number of parameters of the proposed method is the penultimate. The small model size makes our method faster than the others. Accordingly, the run time of the proposed method is also faster than the others.

Finally, as illustrated in Fig.~\ref{Visual}, a comparison of visual quality among various methods is presented. It is evident that the proposed method yields superior visual quality results. A closer analysis of Fig.~\ref{Visual} reveals that our method effectively restores finer texture details, particularly enhancing edge information. This enhancement is notably pronounced when addressing rapid motion-induced edges. In contrast, the impact on large background areas appears relatively modest. This observation aligns with the analysis of experimental data and the visual quality map, underscoring the favorable influence of our proposed network, particularly in scenarios involving intricate textures and fast-motion edge occurrences.

\subsection{Extended application for spatial and angular super resolution of light field}\label{subsec4.4}

Different from conventional cameras, the light field cameras can not only record the light intensity distribution but also record the incident angle of the light rays. Due to the limitations of CCD, the spatial and angular resolution of the light field are contradictory. To obtain high-resolution and high-angle resolution light field, it is necessary to perform space-angle super-resolution. The converted sub-aperture images (SAIs) can be rearranged into a video sequence. The space-angle super-resolution for a light field is actually equivalent to the STSR for a video.

\begin{figure*}[!b]
	\centering
	\includegraphics [width=\linewidth]{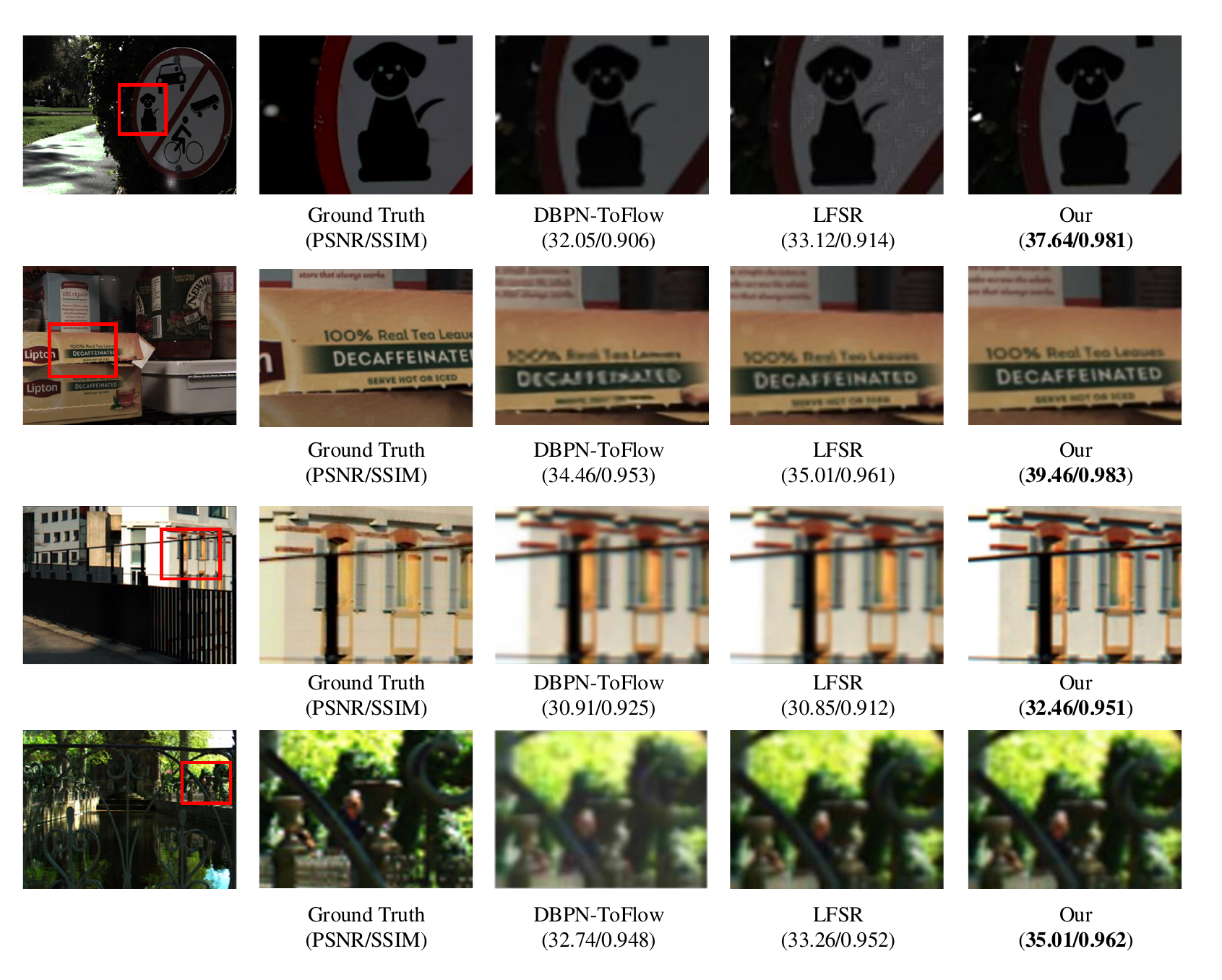}
	\caption{Visual results and objective quality (PSNR/SSIM) comparison on light field images.}
	\label{LF}
\end{figure*}

Due to the nature of light field, there is only a very small disparity between SAIs, therefore we deleted the CFQE module of the proposed Cuboid-Net for the space-angle super-resolution. We adopted two public real-world LFIs datasets, i.e., Stanford Lytro Light Field Archive~\cite{46} and the Light field image data set created by Kalantar~\cite{47}, for train and test. Specifically, the training sets of both of the two datasets were used to train the proposed neural network, and their test sets were also used together for test. 
To verify the effectiveness of  the proposed method, we compared the proposed method with DBPN~\cite{37}+ToFlow~\cite{34} and LFSR~\cite{lfsr}. DBPN~\cite{37}+ToFlow~\cite{34} is a also video super resolution method which was used herein for spatial and angular super resolution of light field. 
LFSR~\cite{lfsr} is a specifically designed CNN-based spatial and angular super resolution approach for light field. The results are also shown in Fig.~\ref{LF}, from which we can see that the proposed method achieves the best, especially in edges.

\section{Conclusion}\label{sec5}

We proposed a Cuboid-Net to directly reconstruct HR and high frame-rate video from LR and low frame-rate video. The proposed network contains four modules, i.e., MBFE module, MBR module, a first stage QE module, and a second stage CFQE module for interpolated frames only. To make full use of the spatial and temporal information, the MBFE uses multi-branch network to achieve the ST-SR of the video from different dimensions. The MBR then fuses the features coming from different branches. The QE is to enhance the quality of the video frames by frame, while the CFQE is used to further improve the image quality of the interpolated frames to tackle the artifacts caused by fast motion. Experimental results showed the effectiveness of the proposed method. 
Although the proposed method is better than the existing methods, the improvement is still limited to some extent, especially for video frames with large motion. In future, we will continue to study the space-time information characteristics in video sequences to further improve the performance. 

\section{Acknowledgments}\label{sec6}

This work was supported in part by the National Natural Science Foundation of China under Grants 62222110, 62172259, the Taishan Scholar Project of Shandong Province (tsqn202103001), the Central Guidance Fund for Local Science and Technology Development of Shandong Province, under Grant YDZX2021002, the Natural Science Foundation of Shandong Province under Grant ZR2020MF139 and ZR2022ZD38.

\end{document}